# Time Preference and Tax Burden Acceptance: Asymmetric Effects on Intertemporal and Contemporaneous Redistribution

Eiji YAMAMURA (Seinan Gakuin Univ)

Fumio OHTAKE (Osaka Univ)

**Abstract**

This paper examines the extent to which individual time preferences are associated with the willingness to accept different tax burdens. The first is an intertemporal redistribution in which a current consumption tax increase is exchanged for a proportional future reduction. The second is a contemporaneous redistribution where the tax burden borne by individuals is transferred directly to those with significantly lower incomes than their own. Using a cross-sectional online survey of approximately 12,000 observations, we found through various regression analyses that higher time preference is negatively associated with acceptance in both domains. Crucially, the negative coefficient is larger in absolute value for contemporaneous redistribution than for the intertemporal one.

# 1. Introduction

Fundamentally, tax policy is a matter of who bears the costs, who reaps the benefits, and at what time. Citizens' willingness to accept tax burdens is shaped not only by self-interest but also by psychological dispositions. One such disposition is the degree to which individuals discount future payoffs relative to present ones. Wade-Benzoni (1999) identifies a self-other trade-off and time delay as critical factors in intergenerational conflict. Extending this framework, Breuer et al. (2022) address this issue by considering a scenario in which an individual can either donate money to another party immediately or delay the donation.[1] Individuals tend to behave more patiently when acting for others than for themselves (Albrecht et al. 2011), indicating that time preference is associated with social distance when making decisions on social issues.

In this study, we raise the question: "Does time preference operate across persons as well as across time?" We compare individuals' time preferences for contemporaneous and intertemporal income redistribution. Prior studies treat time and social discounting as conceptually distinct (Breuer et al. 2022; Kölle & Wenner 2023). However, whether they are governed by a unified psychological mechanism remains unclear. This study tests whether time preference is reflected in contemporaneous decision-making.

In an internet survey, we present respondents with hypothetical intertemporal and contemporaneous redistribution settings. Comparing these settings allows us to isolate the role of time in redistribution decisions. Based on an original internet survey of approximately 12,000 observations, we find that time preference is negatively associated with both contemporaneous and intertemporal redistribution. Importantly, the magnitude is larger for contemporaneous redistribution, indicating an asymmetric effect. This asymmetry is particularly pronounced in the upper tail of the distribution.

---

[1] Social discounting is measured by computing discount factors to evaluate payoffs to others (see also, e.g., Yi et al., 2011 or Charlton et al., 2011: Rachlin and Jones 2008).

Although the analysis does not allow for causal identification, we propose the following interpretation. Time preference is not merely a measure of intertemporal discounting. It also correlates with a broader orientation toward present-focused, self-regarding decision-making (Kim, 2023; Kölle and Lauer, 2024; Falk et al., 2018). This study makes three contributions. First, it tests whether time and social distance are processed through a unified psychological mechanism. Second, it provides novel empirical evidence that time preference is associated with contemporaneous redistribution decisions. Third, by comparing contemporaneous and intertemporal redistribution within the same individuals, it identifies an asymmetric relationship across redistributive domains.

The remainder of the paper is structured as follows. Section 2 reviews the theoretical and empirical background. Section 3 describes the data and measurement strategy. Section 4 states the hypotheses. Section 5 describes the estimation approach. Section 6 reports results. Section 7 discusses the interpretation and validity of the MEL task. Section 8 concludes.

## 2. Literature Review

The empirical measurement of time preferences has been extensively reviewed by Cohen et al. (2020) and Frederick et al. (2002). Subjects are asked to choose between X dollars at an early date or Y dollars at a later date, which is referred to as money earlier or later (MEL) experiments. Cohen et al. (2020) document that MEL responses are driven partly by factors distinct from pure time preference. These include trust in payment institutions and perceived complexity of the choice environment.

Using large-scale cross-country survey data, Falk et al. (2018) found that patience is significantly and positively correlated with savings and investments for education measured by hypothetical questions. They did not explore how patience is correlated with prosocial behaviors although they found the positive association between trust and prosocial behaviors. As closely related to this study, Kim (2023) shows more directly that higher discount factors increase cooperation rates in infinitely repeated games in the laboratory. Kölle and Lauer (2024) provide a comprehensive experimental analysis of cooperation in intertemporal contexts. They find that individual impatience is one of three mechanisms through which delayed benefits reduce cooperation. The other two mechanisms are shifts in beliefs about others' effort and in willingness to conditionally cooperate. However, they also did not examine how time preference is directly associated with prosocial behaviors if other things are equal.

Researchers examine whether present bias is consistent across individual and social contexts (Rong et al. 2019; Kölle & Wenner 2023). Using a longitudinal experiment, they find that participants exhibit present bias when making intertemporal allocation decisions that affect only themselves. However, present bias is not observed when making decisions on behalf of others. This asymmetry implies that the relationship between time preference and prosocial choice is not simply one of discounting social benefits at the same rate as private benefits. Instead, the motivational basis of prosocial decisions differs qualitatively from that of self-regarding decisions. This suggests that time preference will affect contemporaneous redistribution through a channel different from pure discounting.

Charitable giving is time-inconsistent in a specific sense. Social rewards from a giving decision begin accruing at the moment a decision to give has been made. Later, the donor's utility increases again from seeing their donations at work when the gift is transacted. A fundraiser can get more donations if a donor is allowed to decide now to give later (Andreoni and Serra-Garcia 2021). Chopra et al. (2024) extend this analysis by distinguishing consequence-dated and choice-dated prosocial utility. They show that both are pervasive but negatively correlated at the individual level. This suggests a difference in the timing of utility gains derived from prosocial decision-making. Kovarik (2009) found that average gifts decline as payments are delayed, even when both parties face identical delays. This contradicts standard intertemporal utility maximization. It suggests that the act of temporal displacement itself suppresses prosocial impulses. Communicative commitment via promises largely mitigates the time-dampening of trust and trustworthiness over a three-week delay (Ederer and Schneider 2022). This suggests that the relationship between time and prosocial behaviour is contingent upon the availability of commitment technologies.

While the time structure significantly shapes prosocial preferences, its role appears more limited regarding unethical behavior. Bortolotti et al. (2022) demonstrate the persistence of dishonesty, finding that neither delaying the gains from cheating nor providing additional time for reflection reduces the likelihood of dishonest behavior. This suggests that the psychological mechanisms underlying dishonesty may be more robust to time-related interventions whereas positive social utility is sensitive to temporal framing. These results indicate that the relationship between time and self-interested behaviour is not uniform. Patience may promote prosocial cooperation without necessarily suppressing self-interested deception.

While these experimental studies clarify the psychological mechanisms of intertemporal prosociality, their implications extend to broader institutional preference. Alesina and Giuliano (2009) incorporate beliefs about fairness, social mobility and cultural factors, and personal history. Individual preferences for redistribution are also shaped by perceived income differences and social distance (Corneo and Grüner, 2002; Luttmer and Singhal, 2011).

The most directly relevant experimental evidence on time preferences and fiscal attitudes is provided by Tiezzi and Xiao (2016). In an intertemporal experiment with negative externalities, they show that delayed externalities significantly reduce support for Pigouvian taxation. This reduction is driven mainly by the increased perceived complexity of the decision environment rather than by time discounting per se. They argue that providing transparent information about intertemporal trade-offs eliminates the delay effect. Ross et al. (2024) provide contextual evidence from the COVID-19 pandemic. They show that financial vulnerability reduced efficiency concerns while increasing altruism. This underscores the sensitivity of the time preference and prosocial behaviour relationship according to economic conditions.

Yamamura (2025) found having grandchildren is positively associated with supporting an increase in the consumption tax. Based on this finding, he argued that people of the older generation are likely to accept a higher tax burden in order to reduce the fiscal burden on their grandchildren, reflecting intergenerational altruism. However, the well-intended behaviors of prior generations possibly result in unintended outcomes to subsequent generations. It is important to make the intention of past generations transparent in intergenerational resource allocations, which in turn promotes generosity to future others (Bang et al., 2017).

Existing studies treat time preference, social preference, and redistributive attitudes largely as separate domains. However, real-world policy decisions such as redistribution inherently involve both interpersonal and intertemporal trade-offs. Despite this, little is known about whether individuals apply consistent discounting across these dimensions or whether distinct psychological mechanisms govern contemporaneous and intertemporal redistribution.

## 3. Data

### 3.1. Survey Design and Sample

The data were collected through an online survey administered by a Japanese research company. Respondents were drawn from a registered panel. Sampling targeted broad coverage across age groups, genders, and residential prefectures. The analytical sample varies across specifications. The full specification yields 12,273 observations. The alternative specification drops occupation, schooling, and marital status from the independent variables, increasing sample to 18,825 observations.

Our survey elicits time preferences via a standard MEL task (Cohen et al., 2020; Falk et al., 2018; Frederick et al., 2002). We interpret the elicited measure as a composite of time preference and related cognitive-motivational factors. The cognitive uncertainty framework (Enke and Graeber, 2023) further suggests that attenuation bias is more likely in our intertemporal scenario than in the contemporaneous one.

### 3.2. Outcome Variables
### 3.2.1. Maximum Acceptable Tax Rate for Intertemporal Fiscal Adjustment

The intertemporal outcome was elicited through the following scenario: *"Assume that, if left unchanged, the consumption tax rate will reach 40 percent in 30 years. Suppose that each 1 percentage point increase in the current tax rate leads to a 1 percentage point reduction in the future rate. What percentage of tax would you be willing to accept?"* Respondents selected an integer value from 1 to 50 percent. This variable, denoted *Intertemporal*, captures the maximum current tax cost the respondent is willing to bear in exchange for a proportionally smaller future fiscal burden.

The scenario is framed as an intertemporal choice at the individual level. However, a reduction in the consumption tax rate is a public policy outcome. It benefits all members of society simultaneously, not only the individual who accepts a higher current tax. Therefore, the scenario is not a purely private intertemporal trade-off. The personal and the collective dimensions are inseparably bundled. "In addition, the probability of an individual passing away increases with age. Over a 30-year horizon, older respondents face a substantially higher risk of not surviving to benefit from the future tax reduction. This implies that older respondents are less likely to be the direct beneficiaries of their current contributions. Their contributions would thus reduce a tax burden enjoyed primarily by others.

**3.2.2. Maximum Acceptable Tax Rate for Contemporaneous Redistribution**

The contemporaneous outcome was elicited under the following scenario: *"Assume that 80 percent of Japan's population earns less than one-fifth of your income. Further, suppose that the tax paid by you goes directly to those with lower incomes. What percentage of your income would you be willing to pay as tax? Please choose from 1 to 50 percent."* Respondents selected an integer value from 1 to 50 percent. This variable, denoted *Contemporaneous*, measures the respondent's maximum willingness to transfer income to lower-income individuals in the current period.

The scenario contains no intertemporal element. The tax is transferred directly and immediately to lower-income individuals. The respondent therefore receives no future benefit from this payment. This design isolates the prosocial component of redistribution preference. Any willingness to pay reflects concern for others rather than self-interested return."

**3.3. Key Explanatory Variable: Time Preference**

Time preference was elicited through a standard MEL task. Respondents were asked the following question. "What is the minimum amount of money you would need to receive one month from now to forgo receiving 10,000 yen today?" The variable Time Preference is defined as this minimum required future payment, recorded in yen. Higher values indicate stronger present bias. The distribution is right-skewed and includes a lower bound at 10,000 yen and higher bound at 15,000. The higher bound corresponds to an implicit monthly discount rate of 50 percent or higher. Figure 2 illustrates the distribution, exhibiting wide variation in individual time preferences of respondents.

In this study, we use the term *impatience penalty* to refer to the reduction in maximum acceptable tax rates associated with higher time preference. This self-investment framing may reduce the *impatience penalty* relative to a pure intergenerational transfer.

### 3.4. Control Variables and Descriptive Statistics

Table 1 reports descriptive statistics for the full estimation sample. The mean acceptable tax rate is 7.52 percent for the intertemporal scenario and 10.23 percent for the contemporaneous scenario. Both variables have a common range of 1 to 50 percent and a median of 5 percent. The standard deviation is notably larger for contemporaneous redistribution (11.26 versus 7.29). This reflects greater heterogeneity in contemporaneous redistribution preferences.

Figure 1 presents histograms of the maximum acceptable tax rates for both scenarios using identical bin widths. Both distributions are heavily right-skewed, with the mass concentrated between 1 and 10 percent. Specifically, the lowest possible value (1%) was chosen by 31 percent of respondents in the intertemporal redistribution scenario, compared to 24 percent in the contemporaneous one. Interestingly, the proportion of individuals selecting the minimum acceptance rate (1%) is significantly higher for intertemporal redistribution, even though this policy yields future benefits for both society and the individuals themselves. The contemporaneous distribution displays a longer right tail. A non-trivial share of respondents accepts rates above 20 percent for contemporaneous redistribution, whereas intertemporal acceptance rarely exceeds 20 percent. Appendix Figure A1 plots Epanechnikov kernel density estimates (bandwidth = 4) for both variables, confirming greater dispersion in contemporaneous preferences. These distributional differences motivate the use of two-limit Tobit models, which account for the concentration of responses at the lower censoring point.

The mean time preference is 11,676 yen, implying an average monthly premium of approximately 17 percent to defer receipt by one month. Mean household income is 659,000 yen. Mean age is 47.8 years. In the context of intertemporal redistribution set 30 years into the future, respondents will be, on average, nearly 80 years old. The mean number of children is 1.04.

## 4. Hypotheses

Time preference operates through two channels that reduce redistributive behavior. Under the time discounting channel, impatient individuals assign lower weight to future benefits. This reduces willingness to bear current costs for deferred returns. Under the prosocial suppression channel, high time preference is associated with a present-focused, self-regarding disposition. This attenuates concern for others (Kim, 2023; Kölle and Lauer, 2024). Hence, we propose Hypothesis 1.

*Hypothesis 1: Higher time preference reduces the maximum acceptable tax rate in both fiscal domains.*

The motivational basis of prosocial decisions differs qualitatively from that of self-regarding decisions (Kölle and Wenner 2023). This suggests that time preference will affect contemporaneous redistribution in the domain of pure prosocial choice through a channel different from pure discounting. *Hypothesis 1* predicts a negative effect in both domains. The magnitude need not be equal. The intertemporal scenario embeds a partial self-interest motive. This motive is absent in the contemporaneous scenario. Accepting a higher current tax yields a proportional reduction in the respondent's own future tax burden. This self-interested component partially offsets prosocial suppression. In the contemporaneous scenario, the tax transfer accrues exclusively to others. Acceptance therefore depends entirely on prosocial motivation. Therefore, we raise Hypothesis 2.

*Hypothesis 2. The negative effect of time preference is larger in absolute value for contemporaneous redistribution than for intertemporal one.*

In the setting of this study we set a 30-year horizon, older respondents have shorter remaining planning horizons and are unlikely to survive to benefit from the future tax reduction.

This implies that older respondents are less likely to be the direct beneficiaries of their current contributions. This reduces the personal value of future tax reductions. Thus, we predict that the *impatience penalty* intensifies with age. The effect is potentially larger in the intertemporal domain where the self-interest offset is more salient.

Parents have a personal connection with the fiscal environment because their children will inherit it. This increases willingness to bear current costs for future fiscal improvements (Yamamura, 2025). This mechanism operates specifically in the intertemporal domain, where the future fiscal environment is directly affected. We predict that having children attenuates the negative effect of time preference on intertemporal redistribution.

## 5. Methodological Strategy

### 5.1. Two-Limit Tobit Model

Both outcome variables are bounded between 1 and 50 by survey design. As illustrated in Figure 1, censoring is non-trivial in left-censoring at 1 percent. Right-censoring at 50 percent is less frequent but present, particularly for contemporaneous redistribution. We therefore estimate two-limit Tobit models with lower limit 1 and upper limit 50.

The latent willingness to accept tax is specified as:

$y^*_i = \alpha + \beta \, Ln(Time \, Preference_i) + X_i\gamma + \varepsilon_i,$

where the observed outcome is $y_i = 1$ if $y^*_i \leq 1$, $y_i = y^*_i$ if $1 < y^*_i < 50$, and $y_i = 50$ if $y^*_i \geq 50$. We use the log value of *Time Preference* in all regressions. The vector $X_i$ includes Ln(Household Income), Female dummy, Married dummy, Schooling years, Age, and, in the full specification, 18 occupation dummies. In the regression, the full specification includes 18 occupation category dummies and 47 residential prefecture dummies.

### 5.2. Testing Cross-Equation Differences

From *Hypothesis 1*, we **predict** that the coefficient of Time Preference would be negative for both contemporaneous and intertemporal redistribution. To test *Hypothesis 2*, we **examine** the cross-equation difference in the coefficients of Time Preference.

Let $α_1$ denote the coefficient in the intertemporal equation and $β_1$ the coefficient in the contemporaneous equation. We test the null hypothesis that $α_1$ equals $β_1$ against the alternative that $α_1$ is greater than $β_1$. This is equivalent to testing that the intertemporal coefficient is less negative. Results are reported in Table 4 and Appendix Table A1 across four specifications.

*5.3. Interaction Models*

To test the prediction raised earlier, we augment the base Tobit model with interaction terms. We add the interaction *Ln(Time Preference) ×Age* and the interaction *Ln(Time Preference) ×Child*. In each case, the model is estimated separately for the intertemporal and contemporaneous outcomes. The interaction models are reported in Table 3, Panels A and B.

Coefficient of *Ln(Time Preference) ×Age* is the marginal effect of *Ln(Time Preference)* on willingness to accept tax at a given value of *Age*. A negative sign of *Ln(Time Preference) ×Age* implies that the negative effect of time preference strengthens with age. An analogous interpretation applies for the Child interaction. These interaction terms capture heterogeneity of *Ln(Time Preference)* effect.

*5.4. Quantile Regression and OLS Robustness*

Tobit models impose distributional assumptions. To examine whether the asymmetry varies along the distribution of willingness to accept tax, we estimate quantile regressions at the 50th, 75th, and 90th percentiles for both outcomes. As illustrated in Figure 1 and Figure A1 in appendix, both outcome variables exhibit substantial left-censoring: approximately 31 percent of observations are censored at the lower limit of 1 percent for the intertemporal outcome, and 24 percent for the contemporaneous outcome. This concentration near the lower boundary prevents convergence of the quantile regression estimator at or below the 25th percentile, making estimation infeasible at those quantiles. We therefore report estimates at the 50th, 75th, and 90th percentiles only.

Figure 3 plots the quantile coefficients with 95 percent confidence intervals. However, in Figure 3, 95% confidence intervals (CIs) for each equation do not test cross-equation differences. To test the difference, Table 4 reports the cross-equation differences at each quantile where standard errors are calculated based on 500 bootstrap replications.

OLS estimates are presented in Table 5 as a further robustness check. OLS is biased in the presence of censoring. However, it provides an easily interpretable benchmark. The consistency of the OLS and Tobit findings across multiple specifications supports the robustness of our central asymmetry result.

## 6. Estimation Results

### *6.1. Baseline Two-Limit Tobit Estimates (Table 2)*

Table 2 reports baseline Tobit estimates. The key finding is that Ln(Time Preference) is negative and statistically significant at the 1% level in all four columns, consistent with Hypothesis 1. In the full specification, absolute value of the coefficient is 5.171 for the intertemporal outcome (column 1) and 8.233 for the contemporaneous outcome (column 2). The contemporaneous coefficient is thus approximately 59 percent larger in absolute terms. In the alternative specification (columns 3 and 4), similar difference is observed. Both cross-equation differences are significant at the 1 % level, as reported in Appendix Table A1.

The control variable estimates are informative. Ln (Household income) is positively and significantly associated with acceptable tax rate in all columns. This is consistent with lower liquidity constraints and a higher capacity to bear tax burdens. Significant negative sign of the Female dummy is observed at the 1 % level in all results, which shows that females report lower maximum acceptable tax rates. A positive sign of Schooling Years is statistically significant at the 1 % level in all results, which suggests that more educated respondents are more willing to bear redistributive costs by accepting higher tax rates. The coefficient on Age is statistically insignificant with the exception of column (4).

The result of Breuer et al. (2022) and the commitment device interpretation of Andreoni and Serra-Garcia (2021) suggest that the structure of our intertemporal self-investment may partially mitigate prosocial suppression, contributing to the asymmetry we observe.

### *6.2. Interaction with Age: Panel A of Table 3*

Panel A of Table 3 reports estimation results by augmenting the baseline model with the interaction term between Ln(Time Preference) and Age. The sign of the interaction term is consistently negative and statistically significant at the 1% level across all four columns. This indicates that the negative effect of time preference on tax acceptance intensifies with age. Absolute values of the coefficient are 0.219 and 0.145 for intertemporal and contemporaneous specifications in columns 1 and 2, respectively. The value is thus somewhat larger in the intertemporal domain than in the contemporaneous one. Respondents who bear the tax cost today cannot observe the benefits of redistribution immediately. Older individuals are less likely to observe these outcomes within their remaining lifetime. Therefore, the impatience penalty is amplified for older respondents. This is in line with the argument that individuals experience utility from seeing their donations at work (Andreoni & Serra-Garcia, 2021; Kölle & Lauer, 2024).

Sign of *Ln(Time Preference)* becomes positive and significant only in the intertemporal model once the interaction is included. This positive level effect is consistent with the self-interest interpretation. Among the youngest respondents, the prospect of lower future taxes partially offsets the *impatience penalty*. The sign of *Age* is positive and significant in all results. This reflects that older respondents accept higher tax rates if time preference is the lowest level. This is a plausible result if older respondents have clearer views about fiscal sustainability (Okazawa & Takii, 2019).

*6.3. Interaction with Children: Panel B of Table 3*

Panel B of Table 3 reports results of models including the interaction term between *Ln(Time Preference)* and *Child*, where *Child* is the number of respondent's children. The interaction term is positive and significant at the 1 percent level in all four columns. This is consistent with our prediction. Having children substantially attenuates the negative effect of time preference on both intertemporal and contemporaneous tax acceptance.

The interaction coefficient is 6.140 for the intertemporal outcome in the full specification and 5.088 for the contemporaneous outcome. The somewhat larger coefficient in the intertemporal domain is consistent with the prediction that intergenerational altruism specifically raises tolerance for intertemporal redistribution. because the future fiscal environment directly affects children's welfare.

*Child* shows negative sign and statistical significance in all columns. This initially counterintuitive result reflects a compositional effect. Respondents with children have higher household expenditure obligations. This reduces available income for taxation. Once the interaction with time preference is accounted for, the *Child* coefficient captures this fiscal pressure effect if respondents' time preference is the lowest level. The sign of *Ln(Time Preference)* becomes more negative in the presence of the interaction. This reflects the *impatience penalty* for childless respondents.

*6.4. Heterogeneous Effects Across the Distribution: Table 4 and Figure 3*

Table 4 reports quantile regression estimates of the coefficient on Ln(Time Preference) for both outcomes at the 50th, 75th, and 90th percentiles. Figure 3 plots these coefficients with 95 percent confidence intervals.

At the median (q50), the intertemporal and contemporaneous coefficients are nearly identical (−4.996 versus −4.902). The cross-equation difference of −0.095 is statistically indistinguishable from zero. Thus, the asymmetry predicted by Hypothesis 2 is not observed at the median. The median result shows no asymmetry, suggesting that there is no difference between the two domains among respondents who accept only modest tax rates.

At the 75th percentile, in the intertemporal estimation, value of coefficient becomes less negative from −4.96 to negative −1.28. By contrast, the contemporaneous coefficient becomes more negative, from −4.902 to −6.050. This produces a statistically significant difference of 4.774 at the 1 % level. At the 90th percentile, the divergence is even larger. In the intertemporal estimation, value of coefficient is − 3.36 while in the contemporaneous estimation, it becomes − 9.59. The difference is 6.235, while being statistically significant at the 5 % level.

This distributional pattern reveals that the asymmetry is concentrated among respondents who are willing to accept higher tax rates. High-quantile respondents in the intertemporal domain are disproportionately motivated by fiscal sustainability concerns tied to self-interest or civic duty. This makes their intertemporal acceptance relatively insensitive to impatience. By contrast, in the contemporaneous domain, high-quantile respondents become considerably more sensitive to the impatience penalty, This can be explained by lack of their own benefit through burdening taxation in the contemporaneous domain. All in all, the Hypothesis2 is supported for those who accept high tax rates.

The data structure is itself informative. Respondents in the lower tail of the distribution share a common feature. Their maximum acceptable tax rate is driven primarily by a general resistance to taxation rather than by the conceptual difference between intertemporal and contemporaneous redistribution. For such respondents, the motivational channel through which time preference operates is unlikely to be the binding constraint on their response. Whether it operates through discounting of future benefits or suppression of prosocial orientation, neither channel binds when general tax resistance dominates. The absence of meaningful asymmetry at the median is consistent with this interpretation. Even at q50, a substantial share of respondents remains close to the censoring region. The domain difference has limited explanatory power as a motivational factor at lower quantiles.

*6.5. OLS Robustness Checks: Table 5*

Table 5 reports OLS estimates. Ln(Time Preference) is negative and significant in all columns of Table 5. Regarding absolute value of coefficient, for instance in the full specification, the OLS coefficient on Ln(Time Preference) is $-2.52$ for the intertemporal outcome and $-4.85$ for the contemporaneous outcome. As shown in Appendix Table A1, the difference between them is 2.33 while being statistically significant at the 1 % level. Similar results are obtained in alternative specifications. These differences are smaller in magnitude than the differences in the Tobit model. This is expected because OLS attenuates the censored tails of the distribution. Nevertheless, the differences are highly consistent, demonstrating that the observed asymmetry is robust to model choice and specification.

*6.6. Summary of Appendix Table A1*

Appendix Table A1 compares the differences of the coefficient on *Ln(Time Preference)* between Intertemporal and Contemporaneous domains across four specifications: Two-limits Tobit (Full), OLS (Full), Tobit (Restricted controls), and OLS (Restricted controls). The differences range from 1.866 to 3.061, all significant at the 1 % level. Standard errors range between 0.573 and 0.905. The consistency across estimation methods and control sets indicates the robustness of the results to alternative estimation models.

# 7. Discussion

A potential concern in our empirical approach is the use of the MEL task to elicit time preferences. While widely used, MEL responses may reflect not only pure time discounting but also factors such as trust in payment institutions and perceived complexity (Cohen et al. 2020). In our context, if respondents distrust the government's ability to deliver future tax reductions, MEL responses could partly capture institutional skepticism. However, the consistent effect of time preference across both intertemporal and contemporaneous redistribution suggests that our measure captures a broader psychological trait, namely present bias, rather than merely distrust.

The observed difference in coefficient magnitudes between the two domains can be interpreted in terms of cognitive uncertainty (Enke and Graeber 2023). Intertemporal fiscal scenarios, which require evaluating complex trade-offs between current and future tax rates, impose a higher cognitive burden than contemporaneous redistribution decisions. This increased complexity may introduce noise and attenuate estimated effects. Despite this, time preference remains a statistically significant predictor in the intertemporal domain, indicating that its influence is robust even under greater cognitive demands.

Another concern is that MEL tasks capture preferences over monetary payments, whereas tax burdens are experienced as consumption flows. However, in Japan, which is characterized by a rapidly ageing population and high public debt, consumption tax changes are perceived as salient shifts in lifetime purchasing power. It is therefore reasonable to interpret experimentally elicited time preferences as a proxy for how individuals evaluate trade-offs between present and future fiscal burdens.

A further limitation relates to causal identification. Our cross-sectional design does not allow us to establish a causal effect of time preference on redistributive attitudes. Unobserved factors such as political ideology, cognitive ability, or socioeconomic background may influence both. Nonetheless, several features of the results support a structural interpretation. First, the negative association between time preference and tax acceptance is robust across two distinct fiscal domains, multiple estimation methods, and alternative specifications. Second, the interaction patterns are theoretically coherent: the attenuation of the impatience effect among respondents with children is consistent with intergenerational altruism, while its amplification with age aligns with shorter planning horizons.

Taken together, these findings suggest that time preference captures a meaningful and stable psychological disposition shaping redistributive attitudes. While causal identification remains an important avenue for future research, our results provide consistent descriptive evidence that individuals evaluate contemporaneous and intertemporal redistribution through partially distinct behavioral mechanisms.

## 8. Conclusion

This paper documents a systematic asymmetry in the relationship between individual time preferences and fiscal redistribution preferences. Higher time preference reduces the maximum acceptable tax rate for both intertemporal and contemporaneous income redistribution. However, the negative effect is consistently larger for contemporaneous redistribution, and this difference is robust across multiple specifications.

This finding challenges a straightforward application of standard discounting models. While impatient individuals are typically expected to oppose policies with delayed benefits, our results indicate a weaker effect in intertemporal contexts. We interpret this pattern as reflecting the interaction between present bias and self-interest. A present-focused disposition suppresses prosocial orientation, but in intertemporal scenarios this is partially offset by the prospect of future personal gains. This offset is absent in contemporaneous redistribution, generating the observed asymmetry.

Exploratory interaction analyses provide preliminary evidence on two sources of heterogeneity. The *impatience penalty* appears to intensify with age, consistent with shorter planning horizons reducing the personal value of future fiscal consolidation. It appears to attenuate among respondents with children, consistent with intergenerational concern extending the effective planning horizon. These patterns are descriptive and should be treated as tentative. Establishing their causal basis requires designs with exogenous variation in time preferences, age-related planning horizons, or family composition.

Several limitations remain. The cross-sectional design precludes causal inference, and future research using exogenous variation would help identify underlying mechanisms. In addition, further work is needed to assess whether the observed asymmetry reflects cognitive complexity in intertemporal decision-making or a more general behavioral pattern. Cross-national evidence would also clarify the external validity of the findings.

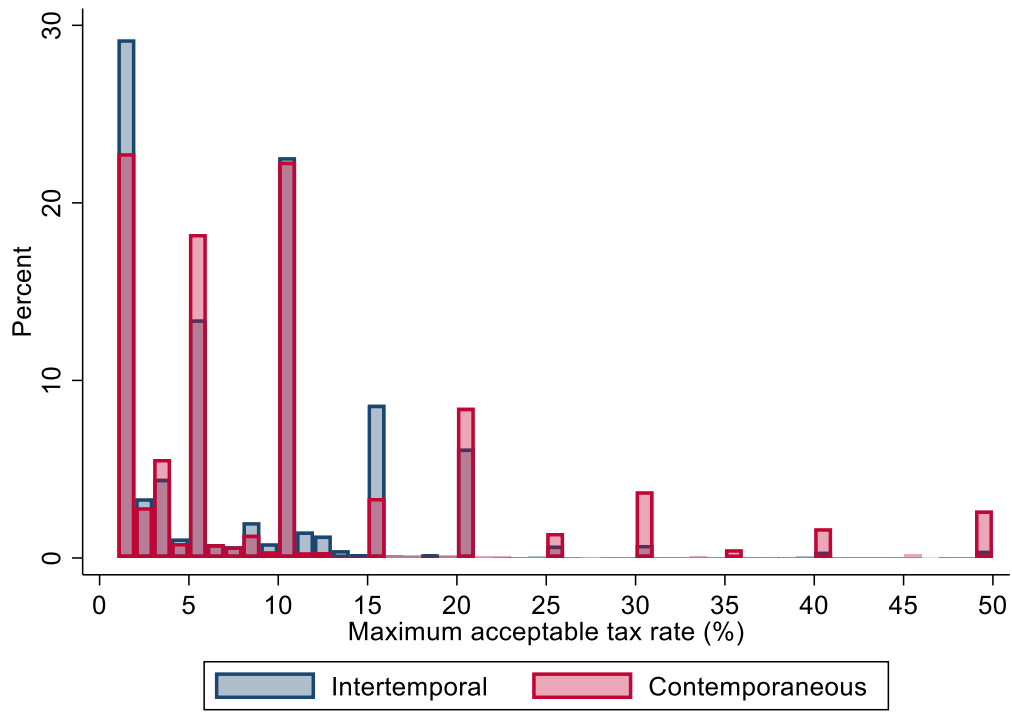

Figure 1. Distribution of Maximum Acceptable Tax Rates for Intertemporal and Contemporaneous Redistribution

*Notes:* The figure presents the distribution of maximum acceptable tax rates for intertemporal fiscal adjustment and contemporaneous income redistribution. Histograms are displayed in percentages using identical bin widths. Intertemporal responses are shown in navy, and contemporaneous responses in cranberry.

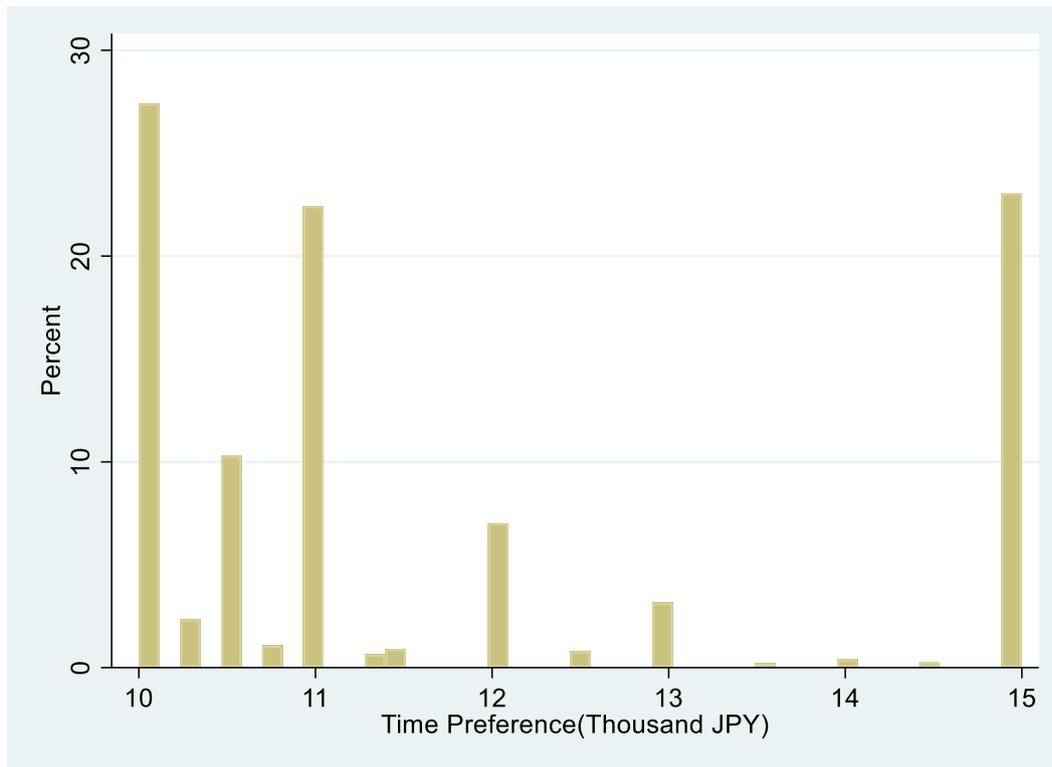

Figure 2. Distribution of Time Preference (Minimum Required Future Payment, Thousand JPY)
Note:The figure reports the distribution of respondents' minimum required future payment (in thousand Japanese yen) to delay receipt of 10,000 JPY by one month. For example, a value of 10 corresponds to 10,000 JPY. Higher values indicate a stronger present bias (higher implicit monthly discount rate). The highest category represents responses of 15,000 JPY or more.

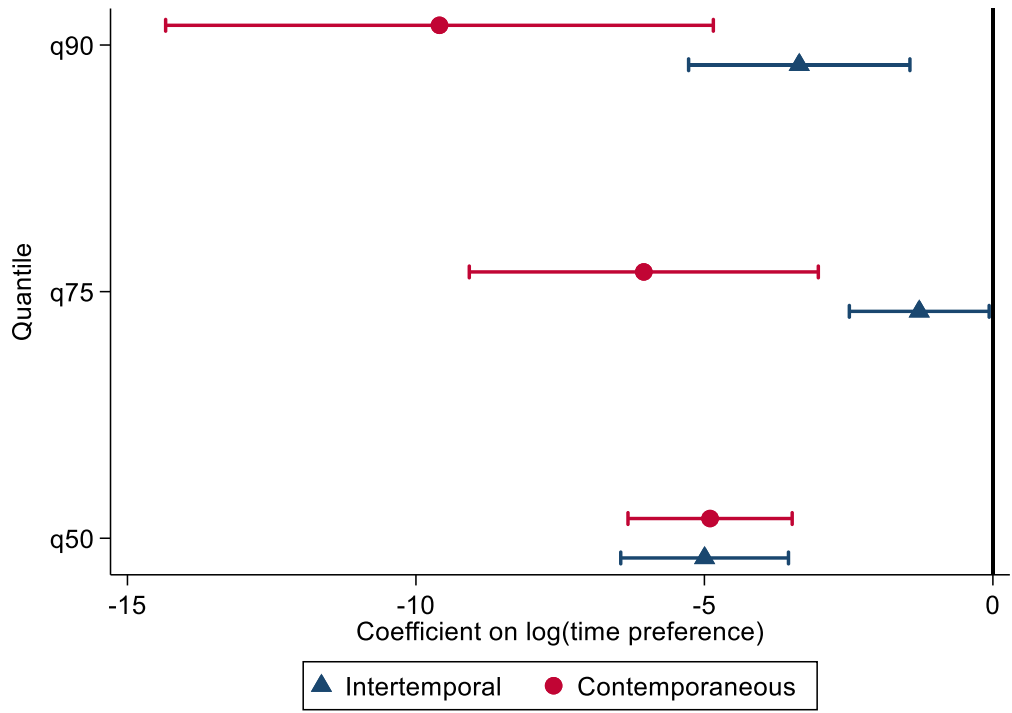

Figure 3. Asymmetric Effects of Time Preference on Maximum Acceptable Tax Rates for Intertemporal and Contemporaneous Redistribution

*Notes:* The figure plots the estimated coefficients of log time preference from quantile regressions (q50, q75, q90) where the dependent variables are the maximum acceptable tax rates for intertemporal fiscal adjustment and contemporaneous income redistribution. Vertical bars indicate 95% confidence intervals (CIs). However, it should be noted that CIs for each equation do not test cross-equation differences. All specifications include the full set of controls. Estimates at the 25th percentile are not reported. Left-censoring at 1 percent accounts for approximately 31 percent (intertemporal) and 24 percent (contemporaneous) of the full sample, rendering sub-median quantile estimates unreliable. The kernel density plots in Appendix Figure A1 confirm the concentration of responses near the lower boundary.

Table 1. Descriptive Statistics

| Variable | Mean | Median | Min | Max | SD |
|---|---|---|---|---|---|
| *Intertemporal* | 7.524 | 5 | 1 | 50 | 7.289 |
| *Contemporaneous* | 10.231 | 5 | 1 | 50 | 11.261 |
| *Time Preference* | 11,675.75 | 11,000 | 10,000 | 15,000 | 1,863.79 |
| *Household Income* | 658.865 | 500 | 50 | 2,300 | 462.360 |
| *Female Dummy* | 0.428 | 0 | 0 | 1 | 0.495 |
| *Married Dummy* | 0.540 | 1 | 0 | 1 | 0.498 |
| *Schooling Years* | 14.834 | 16 | 9 | 18 | 1.980 |
| *Age* | 47.778 | 49 | 18 | 74 | 13.454 |
| *Child* | 1.04 | 0 | 0 | 10 | 1.43 |

Notes

The table reports summary statistics for the estimation sample. Log(Time Preference) is used in regression analysis.

*Intertemporal* and *Contemporaneous* denote the maximum acceptable tax rates (in percent) for intertemporal fiscal adjustment and contemporaneous redistribution, respectively. Time Preference (level) refers to the minimum future amount (in yen) required to delay receipt of ¥10,000 by one month. Household income is measured in thousand yen units. In the empirical analysis, job classification dummies (1–18) and residential prefecture fixed effects (47 categories) are included but not reported.

Table 2. Two-limit Tobit Estimates (Lower = 1, Upper = 50)

|  | (1) Intertemporal (Full) | (2) Contemporaneous (Full) | (3) Intertemporal (Alternative) | (4) Contemporaneous (Alternative) |
|---|---|---|---|---|
| Ln(Time Preference) | -5.171*** (0.935) | -8.233*** (1.229) | -5.882*** (0.767) | -8.415*** (0.854) |
| Ln(Household Income) | 1.277*** (0.225) | 0.884*** (0.254) | 1.480*** (0.179) | 0.930*** (0.182) |
| Female Dummy | -1.216*** (0.252) | -2.334*** (0.363) | -1.297*** (0.209) | -2.833*** (0.239) |
| Married Dummy | 0.401 (0.341) | -0.207 (0.522) | — | — |
| Schooling Years | 0.266*** (0.074) | 0.440*** (0.101) | — | — |
| Age | -0.016 (0.011) | 0.029 (0.020) | 0.001 (0.007) | 0.063*** (0.013) |
| Job dummies | Yes | Yes | No | No |
| Residential dummies | Yes | Yes | Yes | Yes |
| Observations | 12,273 | 12,273 | 18,825 | 18,825 |
| Left-censored | 3,795 | 2,913 | 6,036 | 4,603 |
| Right-censored | 40 | 356 | 62 | 476 |
| LR chi2 | 475.67 | 485.72 | 646.94 | 605.21 |

*Notes:* Two-limit Tobit models with lower limit 1 and upper limit 50. Standard errors (in parentheses) are clustered at the residential prefectures (47 clusters). The full specifications include occupation dummies and residential prefecture fixed effects. The alternative specifications include residential prefecture fixed effects but exclude occupation, schooling, and marital status controls. Coefficients for these controls are not reported. *** $p<0.01$, ** $p<0.05$, * $p<0.10$.

Table 3. Model with Interaction Terms. Two-limit Tobit Estimates (Lower = 1, Upper = 50)

Panel A. Interaction term between Ln(Time Preference ) and Age.

|  | (1) Intertemporal (Full) | (2) Contemporaneous (Full) | (3) Intertemporal (Alternative) | (4) Contemporaneous (Alternative) |
|---|---|---|---|---|
| Ln(Time Preference ) ×Age | -0.219*** (0.045) | -0.145** (0.066) | -0.225*** (0.038) | -0.139*** (0.053) |
| Ln(Time Preference ) | 4.793** (2.165) | -1.560 (3.138) | 4.117** (1.753) | -2.211 (2.476) |
| Age | 2.029*** (0.425) | 1.394** (0.614) | 2.109*** (0.354) | 1.365*** (0.498) |
| Model | (1) Table2 | (2) Table2 | (3) Table2 | (4) Table2 |
| LR chi2 | 498.76 | 490.65 | 682.35 | 605.21 |

Panel B. Interaction term between Ln(Time Preference ) and Child.

|  | (1) Intertemporal (Full) | (2) Contemporaneous (Full) | (3) Intertemporal (Alternative) | (4) Contemporaneous (Alternative) |
|---|---|---|---|---|
| Ln(Time Preference ) ×Child | 6.140*** (0.549) | 5.088*** (0.782) | 3.712*** (0.448) | 2.759*** (0.627) |
| Ln(Time Preference ) | -11.674*** (1.162) | -12.914*** (1.638) | -10.800*** (0.960) | -12.013*** (1.329) |
| Child | -56.921*** (11.00) | -47.233*** (7.366) | -34.437*** (4.448) | -25.697*** (5.908) |
| Model | (1) Table2 | (2) Table2 | (3) Table2 | (4) Table2 |
| Observations | 5,338 | 5,338 | 7,829 | 7,829 |
| Left-censored | 1,559 | 1,719 | 2,301 | 1,733 |
| Right-censored | 14 | 141 | 20 | 183 |
| LR chi2 | 521.08 | 449.93 | 459.71 | 398.28 |

*Notes:* Child is number of respondent's children. Two-limit Tobit models with lower limit 1 and upper limit 50. Standard errors (in parentheses) are clustered at the residential prefectures (47 clusters). Apart from interaction terms, the specifications are equivalent to corresponding columns in Table 2. Coefficients for these controls are not reported. *** p<0.01, ** p<0.05, * p<0.10.

Table 4

Heterogeneous Effects of Time Preference Across Redistributive Domains (Quantile Regressions)

Panel A.

| Quantile | Intertemporal ($\alpha_1$) | SE | Contemporaneous ($\beta_1$) | SE | Difference ($\alpha_1 - \beta_1$) | Boot SE |
|---|---|---|---|---|---|---|
| q50 | -4.996 | (0.742) | -4.902 | (0.726) | -0.095 | (0.836) |
| q75 | -1.277 | (0.618) | -6.050 | (1.543) | 4.774*** | (1.312) |
| q90 | -3.356 | (0.979) | -9.592 | (2.421) | 6.235** | (3.149) |

*Notes:* The table reports quantile regression estimates of the coefficient on log of time preference for maximum acceptable tax rates under intertemporal fiscal adjustment and contemporaneous redistribution. All specifications include income, gender, age, marital status, years of schooling, and occupation dummies. The difference is computed as $\alpha_1$ (Intertemporal) $- \beta_1$ (Contemporaneous). Standard errors are based on 500 bootstrap replications; incomplete replications are discarded.
*** $p<0.01$, ** $p<0.05$, * $p<0.10$.

Table 5. Robustness Checks (OLS): Effect of Time Preference
Dependent variables: Intertemporal (con_tax) and Contemporaneous (alt)

|  | (1) Intertemporal (Full) | (2) Contemporaneous (Full) | (3) Intertemporal (Alternative) | (4) Contemporaneous (Alternative) |
|---|---|---|---|---|
| Ln(Time Preference) | -2.522*** | -4.847*** | -3.036*** | -4.903*** |
|  | (0.758) | (0.894) | (0.610) | (0.582) |
| *Income, Age, Female, Residence* | Yes | Yes | Yes | Yes |
| *Job, Schooling, Married* | Yes | Yes | No | No |
| *R-squared* | 0.033 | 0.038 | 0.029 | 0.029 |
| *Observations (N)* | 12,273 | 12,273 | 18,825 | 18,825 |

Notes: Standard errors (in parentheses) are clustered at the residential prefectures (47 clusters). "Yes" denotes the inclusion of the control variables listed on the left. Statistical significance: *** p < 0.01, ** p < 0.05, * p < 0.10.

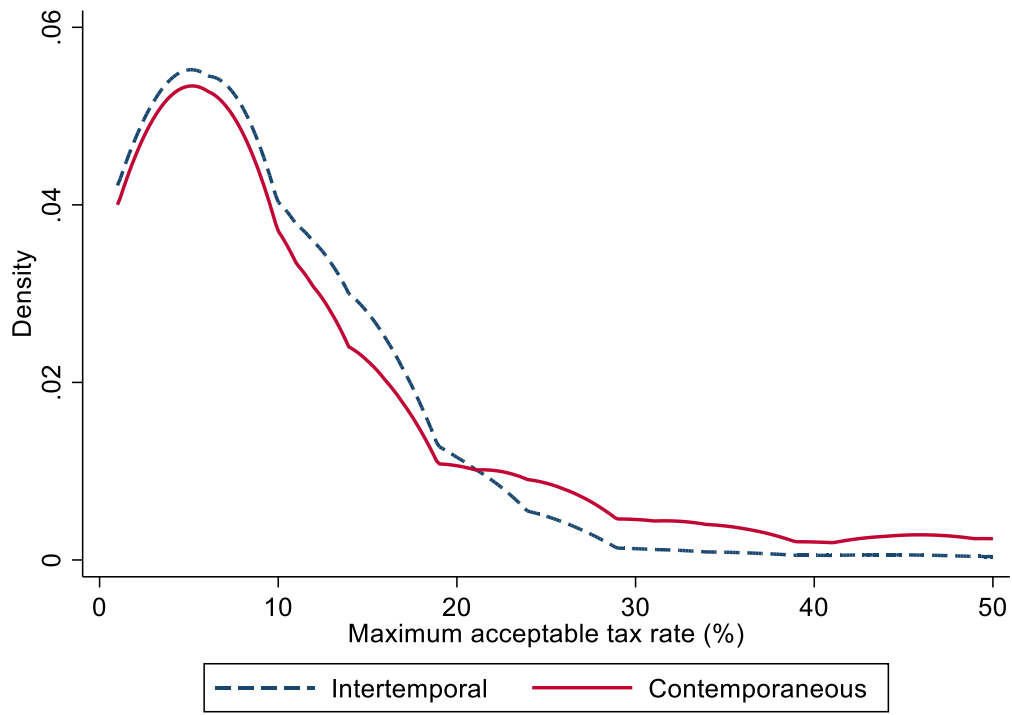

Appendix Figure A1. Kernel Density of Maximum Acceptable Tax Rates
*Notes:* The figure plots Epanechnikov kernel density estimates for Intertemporal (con_tax) and Contemporaneous (alt) redistribution preferences. The bandwidth is set to 4 for both distributions. The support is restricted to the observed range (1–50).

Appendix Table A1
Difference in the Coefficient on log(Time Preference) Across Specifications

| Specification | Difference (Intertemporal − Contemporaneous) | Std. Err. |
|---|---|---|
| Two-limits Tobit (Full) | 3.061*** | (0.905) |
| OLS (Full) | 2.325*** | (0.845) |
| Tobit (Restricted controls) | 2.533*** | (0.661) |
| OLS (Restricted controls) | 1.866*** | (0.573) |

*Notes:* Differences are computed using suest with standard errors clustered at the prefecture level. *** p<0.01, ** p<0.05, * p<0.10.